\documentclass[12pt,twoside]{article}
\usepackage{fleqn,espcrc1,epsfig}

\usepackage{graphicx}
\usepackage[figuresright]{rotating}

\newcommand{\AmS}{{\protect\the\textfont2
  A\kern-.1667em\lower.5ex\hbox{M}\kern-.125emS}}

\title{Nuclear shell effects near the r-process path
\thanks{This work is supported by the Project No. SP056 of the
Research Administration, Kuwait University.}}

\author{M.M. Sharma and A.R. Farhan
\address{Physics Department, Kuwait University,   
        P.O. Box 5969, Safat, Kuwait 13060}}

\begin{document}

% typeset front matter
\maketitle

\begin{abstract}
We have studied the evolution of the shell structure of nuclei near the 
neutron drip line in the Relativistic Hartree-Bogoliubov (RHB) theory
with the vector self-coupling of $\omega$ meson. The experimental 
data on the shell effects about the waiting-point nucleus $^{80}$Zn
are reproduced successfully. It is shown that the shell effects at $N=82$ 
near the r-process path remain strong. A quenching exhibited by the HFB+SkP 
approach is shown not to be compatible with the available data. 

\end{abstract}

\section{INTRODUCTION}

The $N=82$ nuclei at r-process path are assumed to play a significant
role in providing nuclear abundances about $A \sim 130$ \cite{Kratz.93}.
Since nuclei contributing to this peak are extremely neutron-rich
and are not accessible experimentally, there prevail conflicting 
view points \cite{SLHR.94,DHNA.94} on the strength of the 
shell effects near the neutron drip line. In this work, we examine how the 
shell effects evolve with isospin in the region of the astrophysically 
important magic number $N=82$ near the neutron drip line. Using the 
experimental data on the waiting-point nucleus $^{80}$Zn ($N=50$), we 
explore the shell effects near the r-process path in the framework of 
the Relativistic Hartree-Bogoliubov (RHB) theory with the vector 
self-coupling of $\omega$ meson. In this approach, the shell effects
about the stability line have been described successfully \cite{SFM.00}.

\section{THE RMF AND THE RHB APPROACH}
The RMF Lagrangian describes the nucleons as Dirac spinors 
moving in meson fields and is given by \cite{Serot.86}
\begin{eqnarray}
{\cal L}&=& \bar\psi \left( \rlap{/}p - g_\omega\rlap{/}\omega -
g_\rho\rlap{/}\vec\rho\vec\tau - \frac{1}{2}e(1 - \tau_3)\rlap{\,/}A -
g_\sigma\sigma - M_N\right)\psi\nonumber\\
&&+\frac{1}{2}\partial_\mu\sigma\partial^\mu\sigma-U(\sigma)
-\frac{1}{4}\Omega_{\mu\nu}\Omega^{\mu\nu}+ \frac{1}{2}
m^2_\omega\omega_\mu\omega^\mu\\ &&+\frac{1}{2}g_4(\omega_\mu\omega^\mu)^2
-\frac{1}{4}\vec R_{\mu\nu}\vec R^{\mu\nu}+
\frac{1}{2} m^2_\rho\vec\rho_\mu\vec\rho^\mu -\frac{1}{4}F_{\mu\nu}F^{\mu\nu}
\nonumber
\end{eqnarray}
where $M_N$ is the bare nucleon mass and $\psi$ is its Dirac spinor. In 
addition, we have the scalar meson ($\sigma$), isoscalar vector mesons 
($\omega$), isovector vector mesons ($\rho$) and the photons $A^\mu$, 
with the masses $m_\sigma$, $m_\omega$ and $m_\rho$ and the coupling 
constants $g_\sigma$, $g_\omega$, $g_\rho$, respectively. 
A nonlinear scalar potential $U(\sigma) = \frac{1}{2} m^2_\sigma \sigma^2_{} +
\frac{1}{3}g_2\sigma^3_{} + \frac{1}{4}g_3\sigma^4_{}$ for  
$\sigma$-mesons is used for a realistic description of finite nuclei.
Recently, we have also added the non-linear vector self-coupling of 
$\omega$-meson \cite{Bod.91} which is represented by $g_4$. 

For nuclei near a drip line, a coupling to the continuum is required.
A self-consistent  treatment of pairing is also necessary. We employ the 
framework of RHB theory which takes into account both these features. 
The corresponding relativistic Dirac-Hartree-Bogoliubov (RHB) 
equations are obtained as
\begin{equation}
\left(\begin{array}{cc} h & \Delta \\ -\Delta^* & -h^* \end{array}\right)
\left(\begin{array}{r} U \\ V\end{array}\right)_k~=~
E_k\,\left(\begin{array}{r} U \\ V\end{array}\right)_k,
\label{RHB} 
\end{equation}
where $E_k$ are quasiparticle energies and the coefficients $U_k$ and 
$V_k$ are four-dimensional Dirac spinors. The pairing potential 
$\Delta$ in Eq. (\ref{RHB}) is given by
\begin{equation}
\Delta_{ab}~=~\frac{1}{2}\sum_{cd} V^{pp}_{abcd} \kappa_{cd}
\label{gap}
\end{equation}

The RHB calculations are performed by expanding fermionic and bosonic 
wavefunctions in 20 oscillator shells for a spherical configuration. 
We have used the force NL-SH \cite{SNR.93} with the 
non-linear scalar self-coupling and the forces NL-SV1 and 
NL-SV2 with both the scalar and the vector self-couplings. 
The forces NL-SV1 and NL-SV2 have recently been developed
in order to soften the high-density equation of state of nuclear 
matter and are shown to improve the ground-state properties
of nuclei \cite{SFM.00,Sha.00}. 

\section{RESULTS AND DISCUSSION}

In our previous work, it was shown that the existing 
nuclear forces based upon the nonlinear scalar self-coupling of 
$\sigma$-meson exhibit shell effects which were stronger than 
suggested by the experimental data. This problem was solved by 
introducing the nonlinear vector self-coupling of $\omega$-meson 
in the RHB theory \cite{SFM.00}. Having our working basis
established, we explore the shell effects about $^{80}$Zn ($N=50$).
$^{80}$Zn is a waiting point nucleus and here the r-process path comes 
closest to the $\beta$-stability line. We show in Fig. 1 the 
results on $S_{2n}$ values for the Zn isotopes. The kink at 
$N=50$ (Fig 1.a) shows that NL-SH (scalar self-coupling only) 
exhibits shell effects which are much stronger than the 
experimental data. In comparison, the shell gap with NL-SV2 (both the 
scalar and vector self-couplings) \cite{SFM.00} is reduced as compared to 
NL-SH. The shell gap is, however, still larger than the experimental one. 
In Fig.~1(b) we show the $S_{2n}$ values obtained with the force NL-SV1. 
The slope of the kink shows that NL-SV1 is able to reproduce 
well the empirical shell effects in the waiting point region. 
A comparison with the HFB+SkP results shows that 
the shell effects with SkP are strongly quenched. The shell effects were 
shown to be quench with SkP already at the stability line \cite{SFM.00}. 
This behaviour is evidently due to the high effective mass $m^* \sim  1$.

\begin{figure}
\vspace{70pt}
\hspace{80pt}
\epsfig{bbllx=20,bblly=100,bburx=300,bbury=350,width=120pt,file=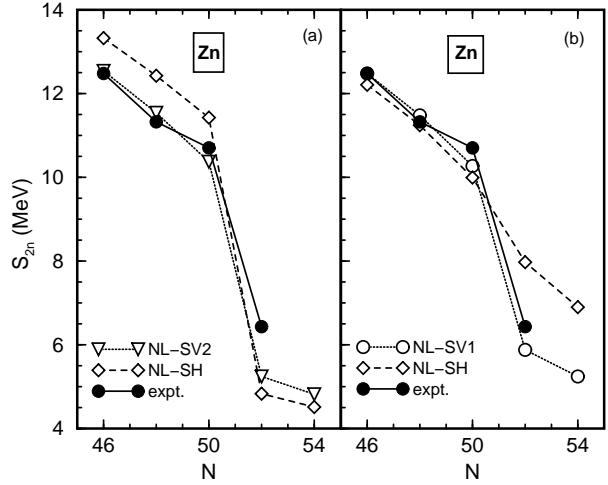}
\vspace{-10pt}
\caption{$S_{2n}$ values for Zn isotopes in the RHB theory.}
\vspace{-10pt}
\end{figure}

With the formalism established in \cite{SFM.00} and tested in
Fig. 1, we extend it to explore the inaccessible region 
of the neutron drip-line about $N=82$. In order to visualize 
the evolution of shell effects, we have chosen the isotopic chains 
of Kr, Sr, Zr and Mo ($Z=36-42$). The results with RHB using
various forces are shown in Fig.~2. For a given force, the shell 
gap at $N=82$ shows a steady decrease in moving from Mo (a) to Kr (d).  
Nuclei become increasingly unbound and a coupling to the continuum 
arises in going to larger neutron to proton ratios. The results with 
the non-linear scalar coupling (NL-SH) show a shell gap at $N=82$, 
which is largest amongst all the forces. Similarly strong shell effects 
with NL-SH were also shown for $N=82$ nuclei near the drip line in 
ref. \cite{SLHR.94} within the BCS pairing. With the force NL-SV2 with 
the vector self-coupling of $\omega$-meson, the shell effects  
are milder as compared to NL-SH for all the chains. The results 
(Fig.~2) with our benchmark force NL-SV1 show that the shell
gap at $N=82$ is reduced as compared to NL-SV2. This is again similar to 
that observed for the Zn isotopes. However, the shell effects with 
NL-SV1 are still stronger as compared to the HFB+SkP results as shown 
in the figure. This is especially true for Mo and Zr isotopes at N=82, 
where the r-process path is assumed to pass through. The quenching 
shown by SkP in this region is as expected. This is consistent with
a strong quenching exhibited by HFB+SkP at the stability line \cite{SFM.00} 
and also near the waiting-point nucleus $^{80}$Zn as shown in Fig.~1(b). 
In comparison, the RHB approach with NL-SV1 shows stronger shell effects 
about the stability line \cite{SFM.00} as well as in the waiting-point 
region at $N=50$, which is consistent with the experimental data.

\begin{figure}
\vspace{150pt}
\hspace{80pt}
\epsfig{bbllx=20,bblly=100,bburx=300,bbury=350,width=120pt,file=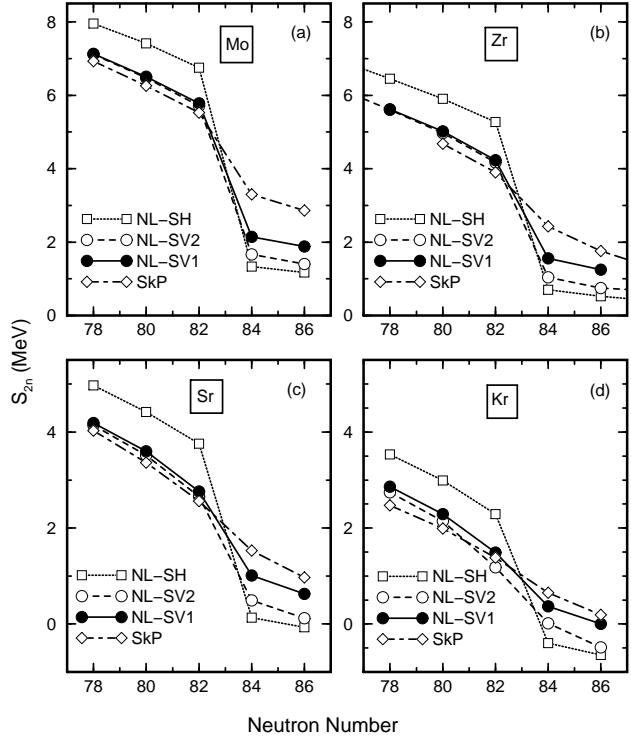}
\vspace{-10pt}
\caption{The RHB results for nuclei near the r-process path.}
\vspace{-10pt}
\end{figure}
The shell effects with NL-SV1 become successively weaker 
as one moves to nuclei with higher isospin such as $^{120}$Sr 
(Fig.~2(c)) and $^{118}$Kr (Fig.~2(d)). It is worth making
a special mention of the Kr nucleus. All the forces 
(except NL-SH) show a complete washing out of the shell effects. 
This stems from the fact that for $^{118}$Kr the Fermi energy 
is very close to the continuum. The shell gap ceases to exist 
and consequently binding energy of an extra neutron is nearly
zero. This implies that the shell effects for $^{118}$Kr ($N=82$) 
are washed out. Such nuclei at the drip line ($\lambda_n \sim 0$) 
are of little interest to the r-process as these nuclei 
can not bind an additional neutron. However, as the r-process path
passes through $S_n \sim 2-4 $ MeV, it is seen (Figs.~2.a and 2.b)
that nuclei which would contribute to the r-process show a 
persistence of stronger shell gaps in contrast to that seen for Kr 
and Sr. Figures 2(a) and 2(b) show that with NL-SV1 the shell effects 
for Mo (Z=42) and Zr (Z=40) nuclei at $N=82$ are much stronger than 
with SkP. For r-process nuclei with $Z > 42$, the shell gaps are 
expected to be even larger. 

\section{CONCLUSION}

We have reproduced the shell effects about the waiting-point nucleus 
$^{80}$Zn with the vector self-coupling of $\omega$-meson in the 
RHB theory. With this basis, it is shown that the shell effects 
near the r-process path about $N=82$ remain strong vis-a-vis a 
quenching exhibited by HFB+SkP. The quenching shown by SkP is
not consistent with the available data. It is noteworthy that 
on the basis of the results with SkP, a quenching has been 
requested for an improved fit to the global r-process abundances 
\cite{Kratz.93}. Since SkP is not able to reproduce the shell 
effects at the stability line and in the waiting-point region, 
an improved fit alone does not necessarily imply a shell quenching 
near the r-process path. Thus, alternative mechanisms for nucleosynthesis 
need to be looked into.

\end{document}